\documentclass[twocolumn,aps,pre,amsmath,amssymb]{revtex4-1}
\usepackage{epsfig}
\usepackage{graphicx}% Include figure files
\usepackage{dcolumn}% Align table columns on decimal point
\usepackage{bm}% bold math
\usepackage[colorlinks=true,dvipdfm]{hyperref}
\usepackage{multirow}
\usepackage{float}

\begin{document}
\title{Nonequilibrium dynamics of the localization-delocalization transition in the non-Hermitian Aubry-Andr\'{e} model}

\author{Liang-Jun Zhai$^{1,2}$}
\author{Guang-Yao Huang$^{3}$}
%\email{guangyaohuang@quanta.org.cn}
\author{Shuai Yin$^{4}$ }
\email{yinsh6@mail.sysu.edu.cn}
\affiliation{$^1$Department of Physics, Nanjing University, Nanjing 210093, China}
\affiliation{$^2$The School of Mathematics and Physics, Jiangsu University of Technology, Changzhou 213001, China}
\affiliation{$^3$Institute for Quantum Information $\&$ State Key Laboratory of High Performance Computing, College of Computer Science and Technology, National University of Defense Technology, Changsha 410073, China}
\affiliation{$^4$School of Physics, Sun Yat-Sen University, Guangzhou 510275, China}
\date{\today}
\begin{abstract}
In this paper, we investigate the driven dynamics of the localization transition in the non-Hermitian Aubry-Andr\'{e} model with the periodic boundary condition. Depending on the strength of the quasi-periodic potential $\lambda$, this model undergoes a localization-delocalization phase transition. 
We find that the localization length $\xi$ satisfies $\xi\sim \varepsilon^{-\nu}$ with $\varepsilon$ being the distance from the critical point and $\nu=1$ being a universal critical exponent independent of the non-Hermitian parameter.
In addition, from the finite-size scaling of the energy gap between the ground state and the first excited state, we determine the dynamic exponent $z$ as $z=2$. 
The critical exponent of the inverse participation ratio (IPR) for the $n$th eigenstate is also determined as $s=0.1197$. 
By changing $\varepsilon$ linearly to cross the critical point, we find that the driven dynamics can be described by the Kibble-Zurek scaling (KZS). Moreover, we show that the KZS with the same set of the exponents can be generalized to the localization phase transitions in the excited states.
\end{abstract}
\maketitle
\section{\label{secIntro}Introduction}
The quasiperiodic system, lying in between the periodic system and the disordered system, exhibits very interesting properties,
such as the topological phase~\cite{verbin2013,Longhi2019,Kraus2012,Zhang2018,Ganeshan2013,Lang2012}, the Anderson localization~\cite{Aubry1980,Sokoloff1981,Sarma1988,Biddle2009,Biddle2011,Luschen2018,Skipetrov2018,Agrawal2020,Sanchez-Palencia1,Sanchez-Palencia2,Sanchez-Palencia3},
the quantized adiabatic pumping~\cite{Kraus2012,Liu2015}, and the cascade of the localization transition~\cite{Goblot2020}.
As a typical model, the Aubry-Andr\'{e} (AA) model~\cite{Aubry1980,Harper1955}, whose on-site potential has an irrational period compared with the lattice period,
has received increasing attention in recent years, partly inspired by its realization in the pseudorandom optical lattice~\cite{Roati2008}, and ultracold atoms~\cite{Billy2008}.
It was shown that the AA model can undergo a localization transition with the change in potential strength~\cite{Aubry1980,Luschen2018,Skipetrov2018,Ganeshan2013},
in contrast to the case of the disordered system which only has the localization phase in one dimension~\cite{Anderson1958,Evers2008,Wei2019}. Moreover, various extensions of AA models have been studied. For example, the energy-dependent mobility edges is found in a generalized AA model with modified quasiperiodic potentials~\cite{Sarma1988} and long-range hopping terms~\cite{Biddle2009,Biddle2011},
and the critical phase lying between the extended and localized phase is found for a generalized AA model with two quasiperiodic modulation parameters~\cite{Liu2015,Thouless1983,Wang2021},
and many-body localization (MBL) has been found in the AA model with the interaction term introduced~\cite{Wang2021,Strkalj2021,Xu2019,Mastropietro2015,zhangsx2018}.

Recently, nonequilibrium dynamics in localized systems has attracted increasing attentions. Lots of exotic properties therein have been discovered. For instance, it is shown that the periodic driving can not only turn the localized eigenstates into extended ones and vice versa~\cite{Molina2014,Bairey2014}, but also bring the system into the topological MBL phase~\cite{Decker2020}.
In addition, a dynamical localization transition can happen when the system is suddenly quenched from the localized phase to the delocalized phase~\cite{Modak2021,Yang2017}.
Furthermore, for the linearly quench across the localization transition point, the driven dynamics can be well described by the Kibble-Zurek scaling (KZS)~\cite{Sinha2019}.

On the other hand, inspired by the experimental progress, the non-Hermitian systems have attracted enormous studies~\cite{Yuto2021,Yao2018,Song2019,Okuma2020,Kawabata2018,Borgnia2020,Longhi2020,Fu2021,Liu2020,Xue2022,
El-Ganainy2018,Bender1998,Mostafazadeh2001,Shen2021,Novitsky2021,SZhang2022,Kunst2018,Bergholtz2021,Xiao2021,zhai20202}. In particular, it has been shown that the interplay of non-Hermiticity and the disordered (or quasiperiodic) potentials can bring intriguing perspectives in the localization phenomena~\cite{Hatano1998,Longhi2019,Longhi20192,Longhi2021,Jiang2019,Jazaeri2001,PWang2019,zhai2020,zhai2021,Tang2021,Cai2021,Cai20212,
chen2022,Goldsheid1998,Liu2021,Liu20212,Liu20201,Guo2021,Xu2022,Zhang2022,Han2022,zhou2022,Suthar2022,Wu2021,Orito2022,Zhou20212,Jiang2021,Cem2022,Markum1999,Kawabata2021,Gong1,Gong2,Gong3}.
For instance, it was shown that for the non-Hermitian AA model with the non-reciprocal hopping the localization transition happens at the same point as the real-complex transition~\cite{Hamizaki2019,zhai2020,Jiang2019,Longhi2019,Liu2021}. For the nonequilibrium dynamics in non-Hermitian localization systems, the periodical driving and the sudden quench have been studied~\cite{Dai2022,Zhou2021,Xu2020,Xu2021}. However, linearly quench dynamics in the localization transition in the non-Hermitian systems is still unexplored.
A natural question is whether the KZS is still applicable in describing the driven dynamics in the localization transition in the non-Hermitian systems.

To answer this question, in the present paper we study the scaling behavior of the driven dynamics of localization transition in the non-Hermitian AA model~\cite{Jiang2019}. The non-Hermiticity of this model is induced by the non-reciprocal hopping.
According to the locations of the critical points of  the localization-delocalization transition and real-complex transition, we classify the eigenstates into three classes. The first class is the ground state, in which the energy spectrum is always real and this type of the state has only the localization transition. The second class corresponds to the eigenstates in which the localization transition and real-complex transition happen at different points. The third class corresponds to the eigenstates in which the localization transition and real-complex transition happen at the same critical point. Then, we study the driven dynamics in this model for various kinds of states. Starting from the deep localized phase and slowly tuning the potential strength across the critical point, we find that the driven dynamics of the localization-delocalization phase transition for different types of initial state can be described by the KZS. Accordingly, we generalize the KZS to localizaion transition in the non-Hermitan quasiperiodic systems.

The remainder of the paper is organized as follows. In Sec.~\ref{secmodel}, the non-Hermitian AA model is introduced, and the static scaling behaviors are presented.
In Sec.~\ref{secKZM}, the driven dynamics of the localiation-delocalization transition is studied. The KZS for various kinds of states is examined by the numerical study.
A summary is given in Sec.~\ref{secSum}.

\section{\label{secmodel}Non-Hermitian AA Model and the static critical properties}
\subsection{The non-Hermitian AA model}

The Hamiltonian of the non-Hermitian AA model reads~\cite{Jiang2019}
\begin{eqnarray}
\label{Eq:model}
H &=& -\sum_{i}^{L}{(J_LC_{j}^+C_{j+1}+J_RC_{j+1}^+C_{j})}\\ \nonumber
&&+2\lambda\cos{[2\pi(\gamma j+\phi)]C_j^+C_j},
\end{eqnarray}
where $C_j^+(C_j)$ is the creation (annihilation) operator of the hard-core boson, $J_L=Je^{-g}$ and $J_R=Je^{g}$ are the asymmetry hopping amplitude in the left and right directions, respectively, $\lambda$ measures the amplitude of the quasiperiodic potential with $\gamma$ therein being an irrational number, and $\phi\in[0,1)$ is a random phase of the potential.
The periodic boundary condition is assumed in the following calculation.
To satisfy the periodic boundary condition, $\gamma$ has to be approximated by a rational number $F_n/F_{n+1}$ where $F_{n+1}=L$ and $F_{n}$ are the Fibonacci numbers. It was shown that all the eigenstates of model~(\ref{Eq:model}) are localized when $\lambda>e^{g}$, whereas all the eigenstates are delocalized when $\lambda<e^{g}$. Thus the critical point for the localization transition is $\lambda_c=e^g$~\cite{Jiang2019,zhai2021}.

\begin{figure}[t]
  \centering
  \includegraphics[width=2.0 in]{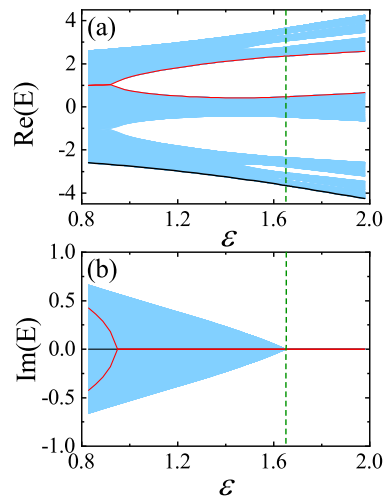}\\
  \caption{(a) Real and (b) imaginary parts of energy spectra of the model Eq.~(\ref{Eq:model}).
  The black curve is the ground state belonging to Class I.
  The red curves denote the eigenstates belonging to Class II with the localization transition and real-complex transition being separated.
  The blues curves denote the third type of eigenstates belonging to Class III with localization transition and real-complex transition happening at the same critical point.
  Here, we choose $g=0.5$, $\phi=0$ and $L=987$ in the calculation.
  The green dashed lines in (a) and (b) label the location of the localization transition critical point.
  }\label{Fig:energy}
\end{figure}

As illustrated in Fig.~\ref{Fig:energy}, the energy spectra of Eq.~(\ref{Eq:model}) are plotted. Here, the eigenstates are arranged in a descending order of the real parts of the eigenenergies. We find that the eigenstates can be classified into three classes. Class I: The first class has the real spectra for all $\lambda$'s. The typical states in this class are the ground states. Class II: The second class corresponds to the states in which the real-complex transition and the localization transition happen at different $\lambda$'s.
The second type eigenstates usually locate at the boundary of the energy bands, i.e., the red curves in Fig.~\ref{Fig:energy}. Class III: The third class is the eigenstate that undergoes a real-complex transition at $\lambda=\lambda_c$ accompanying with the localization transition. We find that most of the excited eigenstates belong to the this class.
\subsection{Static scaling properties for localization transition}
Here we systematically explore the scaling properties of the localization transition for three kinds of spectra. For the ground state in Class I, three quantities can be used to characterize the localization transition. One is the localization length, $\xi$, defined in the localization phase as~\cite{Sinha2019}
\begin{equation}
   \xi = \sqrt{\sum_{n>n_c}^{L} [( n - n_c )^2 ] P_i},
\end{equation}
in which $P_i$ is the probability of the wavefunction at site $i$, $n_c\equiv \sum n P_i$ is the localization center. Near the localization-delocalization transition, for $\varepsilon\equiv \lambda-\lambda_c$, it is expected that $\xi$ satisfies the scaling relation
\begin{equation}
\label{Eq:xiscaling}
   \xi\propto \varepsilon ^{-\nu},
\end{equation}
when $\xi\ll L$.

The second quantity is the inverse participation ratio (IPR) for the $n$th eigenstate, which is defined as~\cite{Bauer1990,Fyodorov1992,Jiang2019}
\begin{equation}
\label{Eq:ipr}
{\rm IPR}_n = \frac{\sum_{j=1}^L||\Psi_n^R(j)\rangle|^4}{\sum_{j=1}^L||\Psi_n^R(j)\rangle|^2},
\end{equation}
where $|\Psi_n^R(j)\rangle$ is the $n$th right eigenvector.
For the extended state with $\lambda<\lambda_c$, the wave function is homogeneously distributed through all sites, and ${\rm IPR}_n$ scales as ${\rm IPR}_n\propto L^{-1}$, whereas for the localization state with $\lambda>\lambda_c$ ${\rm IPR}_n\propto L^0$.
At the localization transition point $\lambda_c$, it is expected that ${\rm IPR}_n$ satisfies a scaling relation
\begin{equation}
\label{Eq:iprscaling}
{\rm IPR}_n\propto L^{-s/\nu},
\end{equation}
with $s$ being its exponent.
When $L\rightarrow\infty$, $s$ can also be determined by
\begin{equation}
\label{Eq:iprscaling2}
{\rm IPR}_n\propto \varepsilon^{s}.
\end{equation}

The third quantity is the energy gap between the first excited state and the ground state. According to the finite-size scaling, this energy gap $\Delta_s$ should scales as
\begin{equation}
\label{Eq:gapscaling}
   \Delta_s\propto L^{-z},
\end{equation}
when $\lambda=\lambda_c$.

For Classes II and III, the energy gap that is relevant to the localization transition cannot be well defined. 
But the scaling relations of the localization length Eq.~(\ref{Eq:xiscaling}) and the ${\rm IPR}_n$ Eqs.~(\ref{Eq:iprscaling}) and (\ref{Eq:iprscaling2}) can also be used to characterize the localization transition in the excited states.
A peculiar situation for Class III is that the real-complex spectra phase transitions happens at the same point as the localization transition. However, the real-complex spectra transition and the localization transition are described by different universality classes.
For the former, as $\lambda$ decreases, a pair of real energy levels converge as $\Delta_r\propto \varepsilon_r^{1/2}$ in which $\varepsilon_r=\lambda-\lambda_r$ is the distance from the real-complex transition point $\lambda_r$, and $\Delta_r$ is the energy difference between these two energy levels for $\varepsilon_r>0$,
whereas for $\varepsilon_r<0$ a pair of complex energy levels generate with the same real part of energy and their imaginary parts of the energies satisfy $\Delta_i\propto |\varepsilon_r|^{1/2}$ in which $\Delta_i$ denotes the imaginary parts of the energy difference. 
Thus, for this transition, the critical exponent $\nu_r z_r=1/2$. This behavior is characterized by the $(0+1)$D Yang-Lee edge singularity~\cite{Yin2017,zhai20203, Fisher1980}.
In contrast, in the following, we will show that for the localization transition, $\nu z=2$.

\begin{figure}[t]
  \centering
  \includegraphics[width=2.0 in]{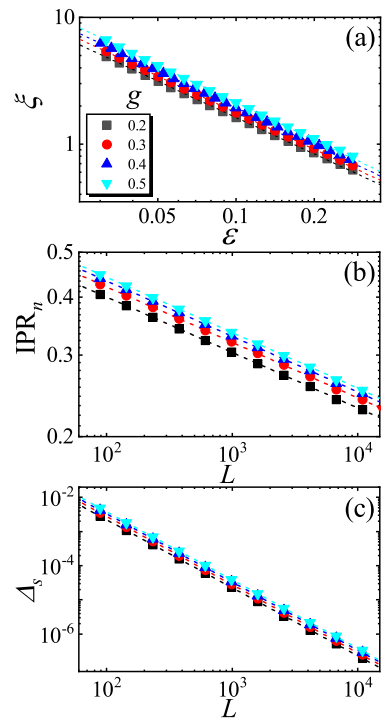}\\
  \caption{Static scaling properties in the ground state. (a) Curves of localization length $\xi$ versus $\varepsilon$ for $L=987$,
  (b) ${\rm IPR}_n$ at $\lambda=\lambda_c$ versus $L$, and (c) Energy gap $\Delta_s$ at $\lambda=\lambda_c$ versus $L$ for different $g$.
  Here, the results are averaged for $100$ choices of $\phi$.
  The power-law fitting yields $\nu=0.9557$, $0.9631$, $0.9719$ and $0.9803$,
  and $s=0.1196$, $0.1197$, $0.1198$, and $0.1198$
  and $z=1.995$, $1.998$, $1.997$, and $1.997$ for $g=0.2$, $0.3$, $0.4$, and $0.5$, respectively.
  }\label{fig:StaticGround}
\end{figure}
\begin{figure}[t]
  \centering
  \includegraphics[width=3.0 in]{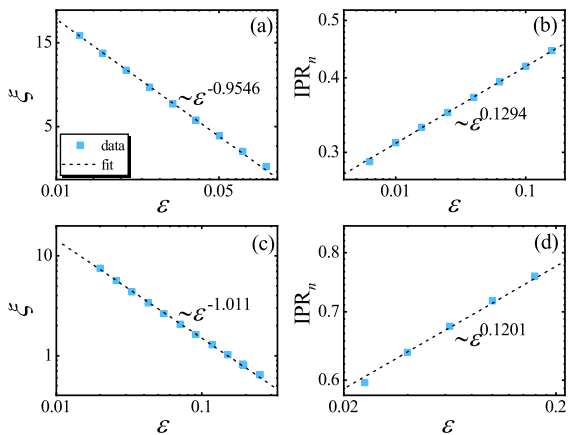}\\
  \caption{Static scaling properties in different kinds of excited states. $\xi$ versus $\varepsilon$ for (a) the $609$th eigenstate (Class II) and (c) the third eigenstate (Class III).
  ${\rm IPR}_n$ versus $\varepsilon$ for (b) the $609$th excited eigenstate (Class II) and (d) the third excited eigenstate (Class III).
 Here, we use $L=2584$ and $g=0.5$.
 The results are averaged for $100$ choices of $\phi$.
  }\label{fig:StaticTypeII}
\end{figure}

In the following, we will numerically examine Eqs.~(\ref{Eq:xiscaling})-(\ref{Eq:gapscaling}) and determine $\nu$, $s$ and $z$.
The numerical results of $\xi$ of the ground state for different $g$'s are shown in Fig.~\ref{fig:StaticGround} (a).
For Class I, Fig.~\ref{fig:StaticGround} (a) confirms the scaling relation of Eq.~(\ref{Eq:xiscaling}) and demonstrates that $\nu=1$.
Moreover, the results for different $g$'s displayed in Fig.~\ref{fig:StaticGround} (a) demonstrate that $\nu$ is a universal exponent independent of $g$.
In addition, Fig.~\ref{fig:StaticGround} (b) shows that ${\rm IPR}_n$ in the localization transition in the ground state satisfies Eq.~(\ref{Eq:iprscaling}) and the averaged critical exponent $s=0.1197$, which is consistent with the result in a non-Hermitian interpolating Aubry-Andr\'{e}-Fibonacci model~\cite{zhai2021}.
Moreover, Fig.~\ref{fig:StaticGround} (c) shows that the energy gap at the critical point satisfies $\Delta_s\propto L^z$ with $z=2$. Note that this value is different from the one obtained for the localization transition in the Hermitian Hamiltonian, where $z=2.37$~\cite{Cestari2011,Sinha2019,Wei2019}.

For Classes II and III, we calculate the localization length and show the results in Figs.~\ref{fig:StaticTypeII}(a) and \ref{fig:StaticTypeII}(c). 
These results demonstrate that the localization length in the excited state also satisfy Eq.~(\ref{Eq:xiscaling}) with $\nu=1$. 
In addition, we calculate the ${\rm IPR}_n$ for the localization transition in the states belonging to Classes II and III. 
Figs.~\ref{fig:StaticTypeII}(b) and \ref{fig:StaticTypeII}(d) show that ${\rm IPR}_n$ at the localization transition for both cases satisfy ${\rm IPR}_n\propto \varepsilon^{0.1294}$ and ${\rm IPR}_n\propto\varepsilon^{0.1201}$, similar to the case of Class I.
These results indicate that the localization transitions for all three classes of states belong to the same universality class. 
In the next section, we will show that the dynamic exponent for both Classes II and III is $z=2$, although it cannot be determined via the gap scaling.

\section{\label{secKZM}Kibble-Zurek Scaling in the Non-Hermitian AA Model}
\subsection{General theory of the KZS}
In usual phase transitions, when the tuning parameter $\varepsilon$ is changed linearly as
\begin{eqnarray}
  \varepsilon &=&-Rt,
\end{eqnarray}
to drive a system cross its critical point, the KZS states that for $|\varepsilon|>R^{1/\nu r}$ with $r=z+1/\nu$ the system can evolve adiabatically since the state has enough time to adjust to the change in the Hamiltonian;
whereas for $|\varepsilon|<R^{1/\nu r}$ the system enters the impulse region and ceases to evolve as a result of the critical slowing down.
However, investigations showed that the assumption that the system stop evolving in the impulse region is quite excessive.
For instance, a finite-time scaling theory demonstrates that in the impulse region the system evolves according to a time scale $\zeta\sim R^{-z/r}$~\cite{Zhong2006,Gong2010,Huang2014}.
Accordingly, for a quantity $Y$, its full scaling form reads
\begin{equation}
\label{Eq:fullfts}
   Y(\varepsilon,R)=R^{y/r}f_Y(tR^{z/r}),
\end{equation}
in which $y$ is the critical exponent of $Y$ and is defined according to the static scaling $Y\propto \varepsilon^y$ when $L\rightarrow\infty$, and $f_Y$ is the scaling function. At the critical point, $t=0$, Eq.~(\ref{Eq:fullfts}) demonstrates that $Y$ can be scaled with $R$ as $Y\propto R^{y/\nu r}$ for $L\rightarrow\infty$.
Equation~(\ref{Eq:fullfts}) was first established in classical phase transitions~\cite{Monaco2002,Anglin1999,Antunes2006}.
In quantum phase transitions, similar scaling forms were also proposed from different perspectives in various systems~\cite{Dziarmaga2005,Damski2007,Chandran2012}.
Recently, the KZS and the full scaling form Eq.~(\ref{Eq:fullfts}) have been generalized into the non-Hermitian Yang-Lee edge singularity~\cite{Yin2017,zhai2018} and the localization transition in the ground state of the Hermitian system~\cite{Sinha2019,Tong2021}.
In the following, we will generalize this scaling form into the localization transition in the non-Hermitian AA model~(\ref{Eq:model}).

\begin{figure}[t]
  \centering
  \includegraphics[width=2.0 in]{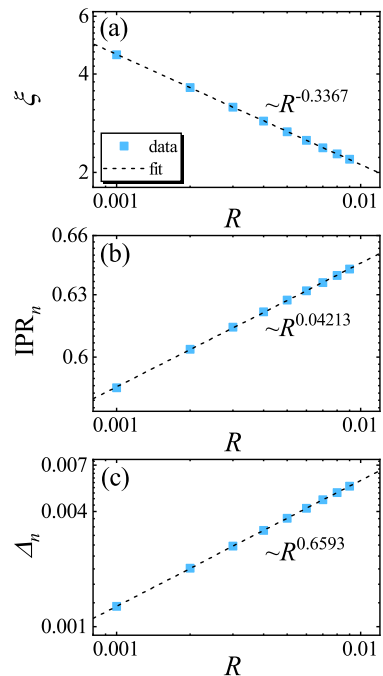}\\
  \caption{(a) ${\xi}$ and (b) ${\rm IPR}_n$ at $t=0$ and (c) ${\Delta}_n$ at $tR^{z/r}=0.5$ as a function of $R$ for the initial ground state.
  Here, we use $g=0.5$ and $L=987$. The results are averaged for $10$ choices of $\phi$.}
  \label{Fig:Fitground}
\end{figure}
\subsection{KZS in states of Class I}
We at first explore the driven dynamics in the states in Class I.
We focus on the driven dynamics in the ground state.
In addition, we set the lattice size $L$ very large and the finite-size effects can be ignored. For the localization length $\xi$, Eq.~(\ref{Eq:fullfts}) converts to
\begin{equation}
 \label{Eq:dynamicScalingXiR}
 \xi(t,R)=R^{-1/r}f_{\xi}(tR^{z/r}),
\end{equation}
in which $r=3$. Besides, according to Eq.~(\ref{Eq:fullfts}), the dynamic scaling form of ${\rm IPR}_n$ is
\begin{equation}
 \label{Eq:dynamicScalingipr}
 {\rm IPR}_n(t,R)=R^{s/r \nu}f_{\rm I}(tR^{z/r}).
\end{equation}
In addition, we define an energy difference $\Delta_n(t)$ between the time-dependent state and the $n$th eigenstate as
\begin{equation}
  \Delta_n(t) \equiv {\rm Re}[\langle \Psi_n^L(t)|H(t)|\Psi_n^R(t)\rangle-E_n],
\end{equation}
where $\langle \Psi_n^L(t)|\equiv\langle \Psi_n^L(0)|e^{iH(t)t}$ and $|\Psi_n^R(t)\rangle\equiv e^{-iH(t)t}|\Psi_n^R(0)\rangle$, $|\Psi_n^{R}(0)\rangle$ and $\langle \Psi_n^L(0)|$ are the right and left eigenstates at the initial time, and $E_n$ is the $n$th eigenenergy for the instantaneous Hamiltonian $H(t)$. The driven dynamics of $\Delta_n$ satisfy
\begin{equation}
  \label{Eq:ScalingEG}
 \Delta_n(t,R)=R^{z/r}f_{\Delta}(tR^{z/r}),
\end{equation}
according to Eq.~(\ref{Eq:fullfts}).

\begin{figure}
  \centering
  \includegraphics[width=3.3 in]{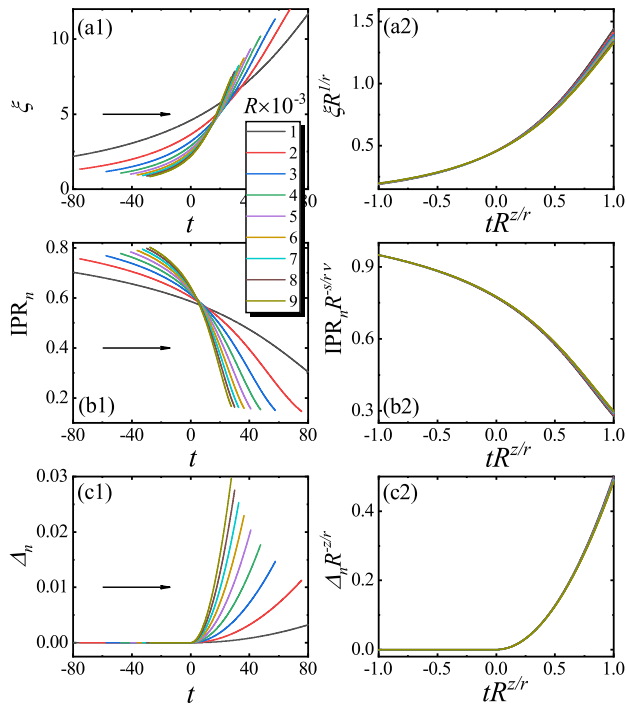}\\
  \caption{Driven dynamics with the initial ground state. The curves of $\xi$ versus $t$ before (a1) and after (a2) rescaling for different $R$'s.
  The curves of ${{\rm IPR}_n}$ versus $t$ before (b1) and after (b2) rescaling for different $R$'s.
  The curves of $\Delta_n$ versus $t$ before (c1) and after (c2) rescaling for different $R$'s.
  Here, we use $g=0.5$ and $L=987$, the results are averaged for 10 choices of $\phi$.
  The arrows in (a1), (b1), and (c1) point to the quench direction.}\label{Fig:dynamicGround}
\end{figure}

\begin{figure}[t]
  \centering
  \includegraphics[width=3.3 in]{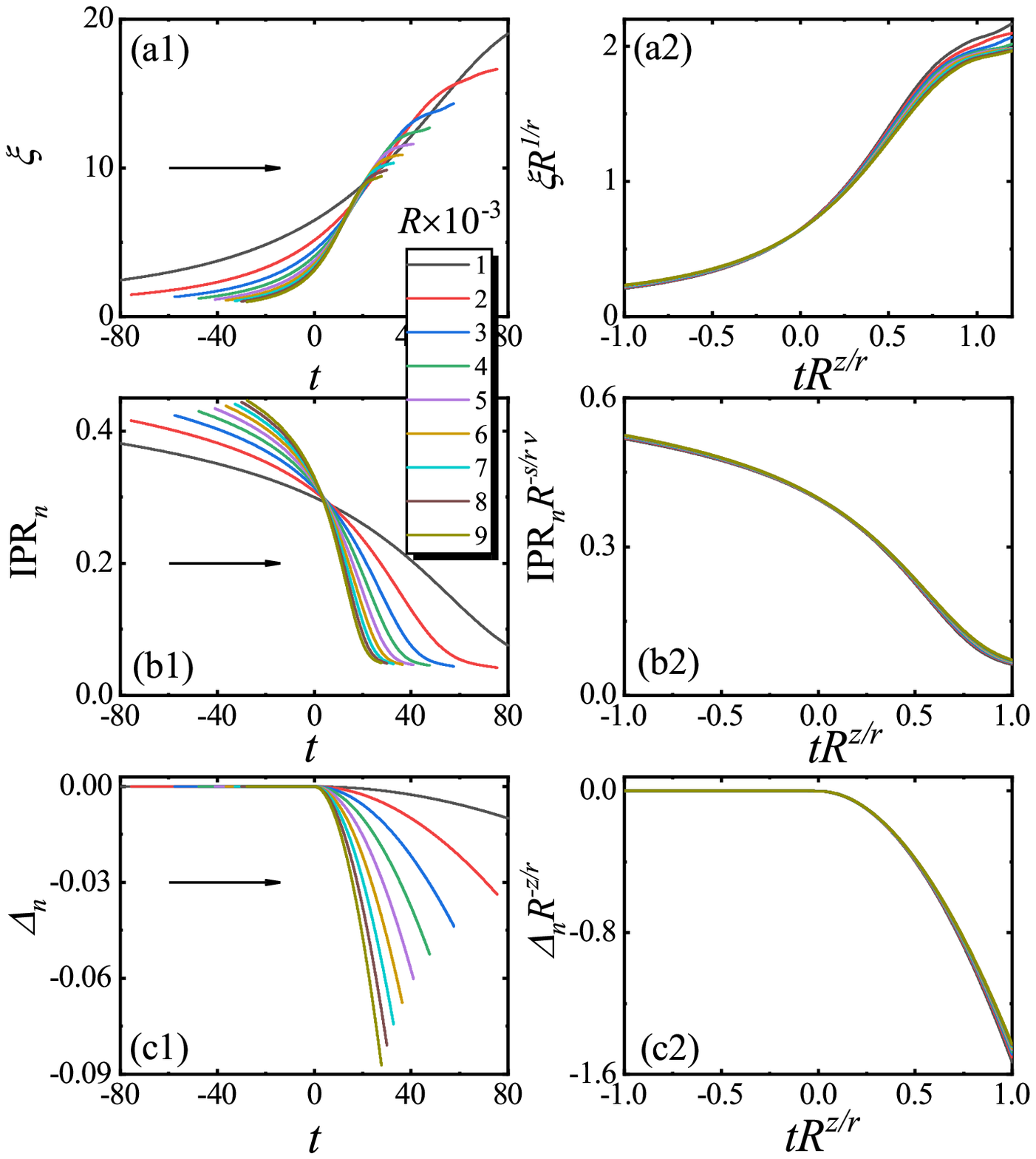}\\
  \caption{Driven dynamics with the initial state belonging to Class II.
  The curves of $\xi$ versus $t$ before (a1) and after (a2) rescaling for different $R$'s.
  The curves of ${{\rm IPR}_n}$ versus $t$ before (b1) and after (b2) rescaling for different $R$'s.
  The curves of $\Delta_n$ versus $t$ before (c1) and after (c2) rescaling for different $R$'s.
  Here, we use $g=0.5$ and $L=987$. We choose $609$th excited state as the initial state in the figure. The results are averaged for $10$ choices of $\phi$.
  The arrows in (a1), (b1) and (c1) point to the quench direction.
}\label{fig:dynamicTypeII}
\end{figure}

\begin{figure}
  \centering
  \includegraphics[width=3.3 in]{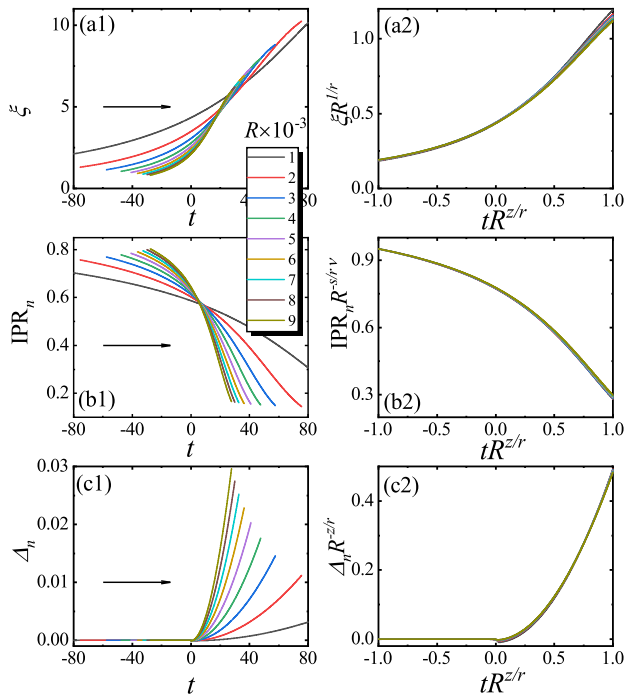}\\
  \caption{Driven dynamics with the initial state belonging to Class III.
  The curves of $\xi$ versus $t$ before (a1) and after (a2) rescaling for different $R$'s.
  The curves of ${{\rm IPR}_n}$ versus $t$ before (b1) and after (b2) rescaling for different $R$'s.
  The curves of $\Delta_n$ versus $t$ before (c1) and after (c2) rescaling for different $R$'s.
  Here, we use $g=0.5$ and $L=987$. The third excited state is taken as the initial state in the figure. 
  The results are averaged for $10$ choices of $\phi$.
  The arrows in (a1), (b1) and (c1) point to the quench direction.
}\label{fig:dynamicTypeIII}
\end{figure}

Then, we numerically examine Eqs.~(\ref{Eq:dynamicScalingXiR}), (\ref{Eq:dynamicScalingipr}) and (\ref{Eq:ScalingEG}) for the states of Class I.
First, we show in Fig.~\ref{Fig:Fitground} (a) that at the localization transition point $\xi$ satisfies $\xi\propto R^{-0.3367}$ in the ground state.
The critical exponent is close to $-1/r$.
Furthermore, by rescaling $\xi$ and $t$ as $t R^{z/r}$, one finds that the rescaled curves collapse onto each other for the ground state as shown in Figs.~\ref{Fig:dynamicGround}(a1) and ~\ref{Fig:dynamicGround}(a2), confirming Eq.~(\ref{Eq:dynamicScalingXiR}).
Second, Fig~\ref{Fig:Fitground} (b) shows that at the critical point ${\rm IPR}_n$ obeys ${\rm IPR}_n\propto R^{0.04213}$ for the ground state.
Then, by rescaling ${\rm IPR}_n$ and $t$ as ${\rm IPR}_n R^{-s/r\nu}$ and $t R^{z/r}$, respectively, one finds that the rescaled curves collapse onto each other for the ground state as shown in Fig.~\ref{Fig:dynamicGround}(b1) and \ref{Fig:dynamicGround}(b2), confirming Eq.~(\ref{Eq:dynamicScalingipr}).
Third, Fig~\ref{Fig:Fitground} (c) shows that $\Delta_n$ obeys $\Delta_n \propto R^{0.6593}$ at $t R^{z/r}=0.5$ for the ground state.
Then, by rescaling $\Delta_n$ and $t$ as $\Delta R^{-z/r}$ and $t R^{z/r}$, respectively, one finds that the rescaled curves collapse onto each other for the ground state as shown in Figs.~\ref{Fig:dynamicGround}(c1) and \ref{Fig:dynamicGround}(c2), confirming Eq.~(\ref{Eq:ScalingEG}).

\subsection{KZS in states of Class II and Class III}

In this section, we will show that Eqs.~(\ref{Eq:dynamicScalingXiR}), (\ref{Eq:dynamicScalingipr}), and (\ref{Eq:ScalingEG}) are also applicable for localization phase transitions in the states beloging to Classes II and III.
For the Class II state, the real-complex transition happens at smaller value of $\varepsilon$, comparing with the localization transition point.
Accordingly, the localization transition happens at the the real-spectra region.
Figures~\ref{fig:dynamicTypeII}(a1), \ref{fig:dynamicTypeII}(b1) and \ref{fig:dynamicTypeII}(c1) show the evolution of $\xi$, ${\rm IPR}_n$ and $\Delta_n$, respectively, for the $609$th excited state.
After rescaling according to Eqs.~(\ref{Eq:dynamicScalingXiR}), (\ref{Eq:dynamicScalingipr}), and (\ref{Eq:ScalingEG}) with the same set of the critical exponents, we find the rescaled curves collapse onto each other as shown in Figs.~\ref{fig:dynamicTypeII}(a2), \ref{fig:dynamicTypeII}(b2) and \ref{fig:dynamicTypeII}(c2).
These results confirm that the rescaling functions Eqs.~(\ref{Eq:dynamicScalingXiR}), (\ref{Eq:dynamicScalingipr}) and (\ref{Eq:ScalingEG}) are applicable for Class II eigenstates.
In particular, for $\Delta_n$, we find that it decreases as $t$ increases, different from the case of the initial ground state, as shown in Fig.~\ref{Fig:dynamicGround} (c1).
The reason is that when the initial state is the ground state, the energy can only increase under external driving. 
In contrast, when the initial state is in the excited state, the energy can spread to both higher and lower states. 
In the present $609$th excited state, the lower-energy excitation dominates, as shown in Figs.~\ref{fig:dynamicTypeII}(c1) and \ref{fig:dynamicTypeII}(c2).

For the Class III state, we choose the third excited state as an example. We calculate the evolution of $\xi$, ${\rm IPR}_n$, and $\Delta_n$ and show the results in Fig.~\ref{fig:dynamicTypeIII}. 
We find that the rescaled curves, according to Eqs.~(\ref{Eq:dynamicScalingXiR}), (\ref{Eq:dynamicScalingipr}), and (\ref{Eq:ScalingEG}) with the same set of the critical exponents, collapse onto each other. 
These results confirm that the scaling functions Eqs.~(\ref{Eq:dynamicScalingXiR}), (\ref{Eq:dynamicScalingipr}), and (\ref{Eq:ScalingEG}) are applicable for Class III eigenstates.

For the Class III state, the real-complex transition happens at the same point as the localization transition. Since these two kinds of phase transitions belong to different universality classes, a nature question is whether the real-complex transition affects the universal scaling behavior.
To answer this question, we compare the relevant exponents for these two transitions. For the scaling of the correlation length $\xi\sim R^{-1/r}$, the exponent $-1/r$ is always $-1/3$ for both the localization transition and the real-complex transition. 
So the scaling of the correlation length satisfies the same scaling form for the real-complex transition and the localization transition.
Accordingly, from the scaling of the correlation length, one cannot distinguish the contribution of the real-complex transition from that of the localization transition. 
For the energy scaling $\Delta_n\sim R^{z/r}$, the exponent $z/r$ is $2/3$ for the localization transition but is $1/3$ for the real-complex transition.
For the small driving rate, it seems that the contribution from the real-complex transition can dominate. 
However, there are only two relevant states for the real-complex transition. For instance, for the third state, only the fourth state is relevant. 
The energy gap of these two states vanishes at the transition point and becomes complex valued in the other side of the transition point. 
However, for the localization transition, there are lots of states that can be occupied under the external driving. 
Thus, the contribution from the localization transition will dominate. This argument is supported by the scaling of $\xi$ shown in Fig.~\ref{fig:dynamicTypeIII} (c).

\section{\label{secSum}Summary}
 To summarize, we have studied the driven dynamics of the non-Hermitian AA model. 
 We have first determined the static exponent $\nu$, $s$ and $z$ by investigating the static behavior of $\xi$, ${\rm IPR}_n$ and $\Delta_s$, respectively. 
 Then we have studied the driven dynamics of the localization-delocalization transitions for three classes of states. 
 We have found that the driven dynamics in all of these states can be described by the KZS with the same set of critical exponents. 
 Our paper generalizes the KZS to the localization transition in both the ground state and the excited states in non-Hermitian systems.

\section*{Acknowledgments}
We thank A. Sinha for helpful discussions.
L.-J.Z. was supported by China Postdoctoral Science Foundation (Grant No. 2021M691535) and the National Natural Science Foundation of China (Grant No. 11704161).
S. Y. was supported by the National Natural Science Foundation of China (Grant No. 41030090).

\end{document}